\documentclass[epjCONF,columns]{svjour} 
\usepackage{graphicx}
\usepackage[varg]{txfonts} 
\usepackage[latin1]{inputenc}
\session-title{2011 Hadron Collider Physics symposium (HCP-2011)}
\begin{document}
\title{W/Z properties and V+jets at the Tevatron}
\author{Darren~D.~Price\thanks{\email{dprice@fnal.gov}} (on behalf of the D\O\ and CDF collaborations)}
\institute{Department of Physics, Indiana University, Bloomington, IN 47405, USA.}
\abstract{
We present a summary of recent measurements of $W$ and $Z$ properties and $W/Z$ production in association with jets in $p\bar{p}$
collisions at $\sqrt{s}=1.96$~TeV with the CDF and D\O\ detectors. 
Latest measurements of $Z/\gamma^*$ transverse momentum and are presented along with new measurements of the angular distributions of final state electrons
from Drell Yan events as a way to probe $Z$ boson production mechanisms.
The mass dependence of the forward-backward asymmetry in $p\bar{p} \rightarrow Z/\gamma^{*} \rightarrow e^+e^-$ interactions is measured, the effective weak mixing angle extracted,
and the most precise direct measurement of the vector and axial-vector couplings of $u$ and $d$ quarks to the $Z$ boson presented.
New measurements of jets produced in association with $Z$ and $W$ bosons for inclusive, beauty and charm jets are also discussed.
} 
\maketitle
\section{Introduction}
\label{intro}

Precision Tevatron measurements of $W$ and $Z$ boson properties and $W/Z$ boson production in association with jets continue to provide rich
legacy for the understanding of Standard Model processes, both for future precision measurements that are subject to these processes as significant
backgrounds and for searches of physics beyond the Standard Model that have the same final state signatures.

This contribution presents the latest measurements of $W$ and $Z$ bosons produced inclusively and in association with jets,
representing world-leading precision measurements of electroweak parameters and tests of perturbative QCD (pQCD) theory 
across a wide kinematic range, for high jet multiplicities and in events with heavy-flavour jet components.

\section{{\boldmath $Z/\gamma^*$} transverse momentum}

Recent measurement of the $Z/\gamma^*$ transverse momentum\,\cite{d0z} made extensive comparison of corrected data against theoretical predictions
that had varied success in describing the data. The measurement was dominated at low $p_T$ by uncertainties arising from corrections for
experimental resolution and efficiency that limited the precision of studies in this region. A new observable, $\phi^*_\eta$ (see Ref.\cite{phistar} for definition),
highly correlated with transverse momentum, was proposed to allow high precision study of this low $p_T$ region due to its dependence only on the lepton directions
that are experimentally measured with much higher precision than lepton momenta. 

\begin{figure*}[htbp]
  \begin{center}
  \includegraphics[width=0.60\textwidth]{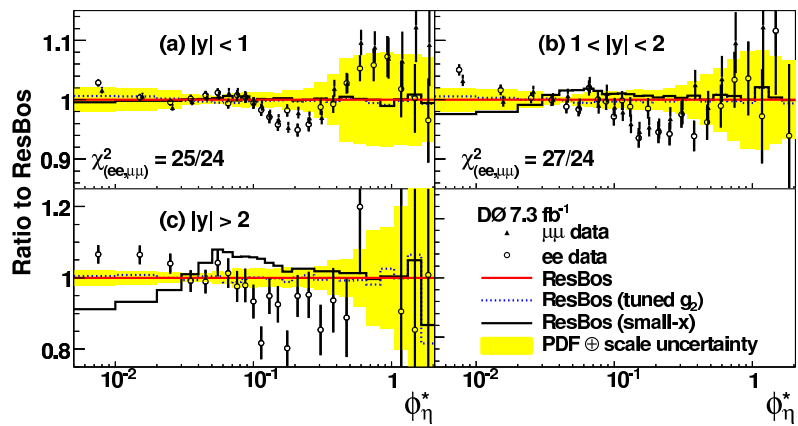}
  \caption{Ratio of the corrected distributions of $(1/\sigma)\times(d\sigma/d\phi^*_{\eta})$ to \textsc{ResBos} for three rapidity intervals.
    For the first two rapidity intervals, both dielectron and dimuon data are presented. Comparison is also made to a variety of \textsc{ResBos}
    models. Differences between data and predictions are observed, and in particular \textsc{ResBos} with small-$x$ broadening is disfavoured.
  }
  \label{fig:phistar}
  \end{center}
\end{figure*}

Using 7.3~fb$^{-1}$ of integrated luminosity, the normalised
differential production cross section of $Z/\gamma^*\to\ell^+\ell^-$ as a function of $\phi^*_\eta$ was measured\,\cite{phistar} by D\O. Figure~\ref{fig:phistar} shows
a comparison of the corrected data compared to Monte Carlo (MC) predictions from \textsc{ResBos}. While the general shape of the distribution is reproduced,
the precision of the data reveal some some areas of significant discrepancy in modelling. The width of the $\phi^*_\eta$ distribution becomes
narrower with increasing rapidity faster in data than predicted by \textsc{ResBos} and in particular the small-$x$ broadening model is seen to have
poor agreement with the data.

\section{{\boldmath $Z/\gamma^*$} angular distributions}

CDF measured\,\cite{DrellYan} the angular distributions of final state electrons from $p\bar{p} \rightarrow Z/\gamma^{*} \rightarrow e^+e^-$ interactions in the Collins-Soper frame
for the invariant mass interval $66<M_{e+e-}<116$~GeV using 2.1~fb$^{-1}$ of integrated luminosity in order to extract measurement of the angular coefficients $A_i$ as
a function of the $Z$ transverse momentum. 

From pQCD, the Lam-Tung relation suggests that $A_0$ and $A_2$ have a specific dependence on $p_T$ (dependent on the contribution
from quark-antiquark annihilation and Compton scattering processes) but should be equal to each other up to corrections of order $\alpha^2_s$.

A strong $p_T$ dependence was observed with $A_{0,2}$. The results were compared to a variety of Monte Carlo predictions and are used to assess the relative contributions
of Compton and annihilation processes to $Z/\gamma^{*}$ production, highlighting that at low $p_T$ production dominantly occurs via annihilation processes, with Compton scattering
playing an increasingly large role at higher $p_T$. 
The average $A_0-A_2$ value across the $p_T$ range studied was $0.02\pm 0.02$.
As the Lam-Tung relation is only valid for spin-1 gluons, this result confirms the vector nature of the gluon.
By contrast to $A_{0,2}$, $A_3$ and $A_4$ are expected (and confirmed by data) to have a value independent of $p_T$, and from $A_4$ a (theory-dependent) 
determination of the weak mixing angle is extracted to be $\sin^2\theta_W = 0.2329 \pm 0.0012$.

\section{{\boldmath $Z/\gamma^*$} forward-backward asymmetry}

\begin{figure}[htbp]
  \begin{center}
  \includegraphics[width=0.4\textwidth]{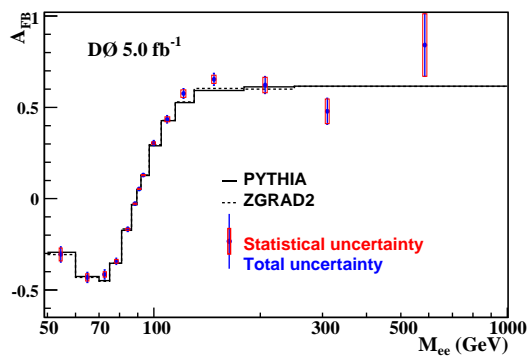}
  \caption{Unfolded forward-backward asymmetry measurement (points) compared with predictions from \textsc{pythia} (solid line) and \textsc{zgrad2} (dashed line).
    No significant departure from expectation is observed across the range $50<M_{e+e-}<1000$~GeV studied.}
  \label{fig:AFB}
  \end{center}
\end{figure}

The presence of both vector and axial vector couplings in $Z/\gamma^*\to\ell^+\ell^-$ production gives rise to an asymmetry in the polar angle $\theta^*$ of the negatively-charged lepton
relative to the incoming quark direction in the $Z/\gamma^*$ rest frame, with events having this lepton $\cos\theta^*>0$ being classified as `forward' and those with 
$\cos\theta^*<0$ being classified as `backward'.
The mass dependence of this forward-backward charge asymmetry $A_{FB}$ was studied\,\cite{ZuZd} by D\O\ using $p\bar{p} \rightarrow Z/\gamma^{*} \rightarrow e^+e^-$ data corresponding to
an integrated luminosity of 5.0~fb$^{-1}$. Any deviation from predictions at high mass could be due to the presence of new physics effects. No such discrepancy was observed. 
The unfolded asymmetry measurement as a function of dilepton invariant mass is shown in Figure~\ref{fig:AFB}. 

In the vicinity of the $Z$ pole, $A_{FB}$ is sensitive to the charged lepton effective mixing angle. This angle is extracted from the detector-level data before corrections 
by comparing the measured background-subtracted distribution with templates simulated with \textsc{pythia} and \textsc{zgrad2} using a variety of $\sin^2\theta^\ell_\textrm{eff}$ inputs.
This allows for a precise measurement of the mixing angle without introducing systematic uncertainties in the unfolding process. Using events in the range
$70<M_{e+e-}<130$~GeV a measurement of $\sin^2\theta^\ell_\textrm{eff} = 0.2304 \pm 0.0008~\textrm{(stat.)} \pm 0.0006~\textrm{(syst.)}$ is extracted.
\begin{figure}[htbp]
  \begin{center}
  \includegraphics[width=0.34\textwidth]{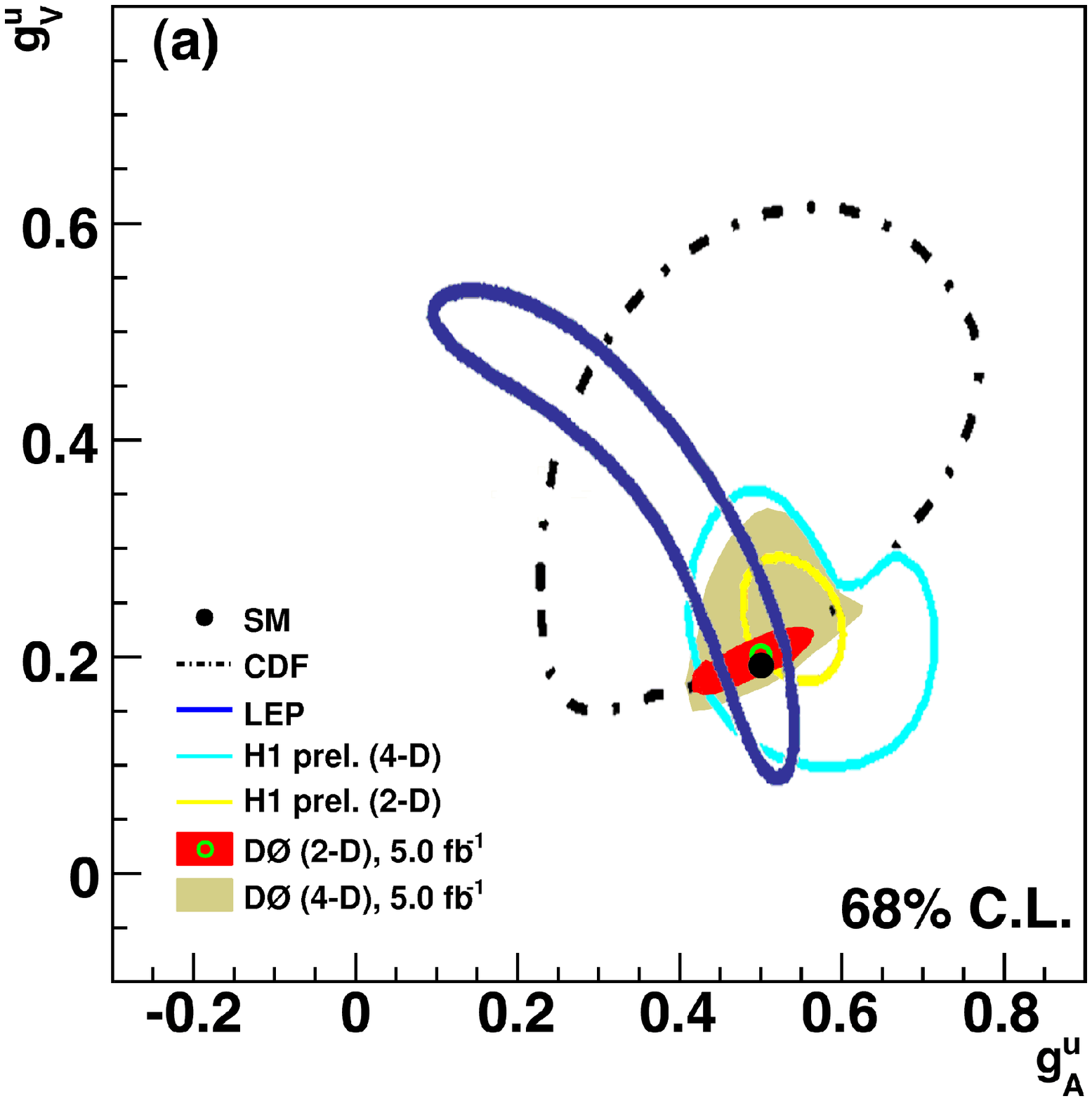}
  \includegraphics[width=0.34\textwidth]{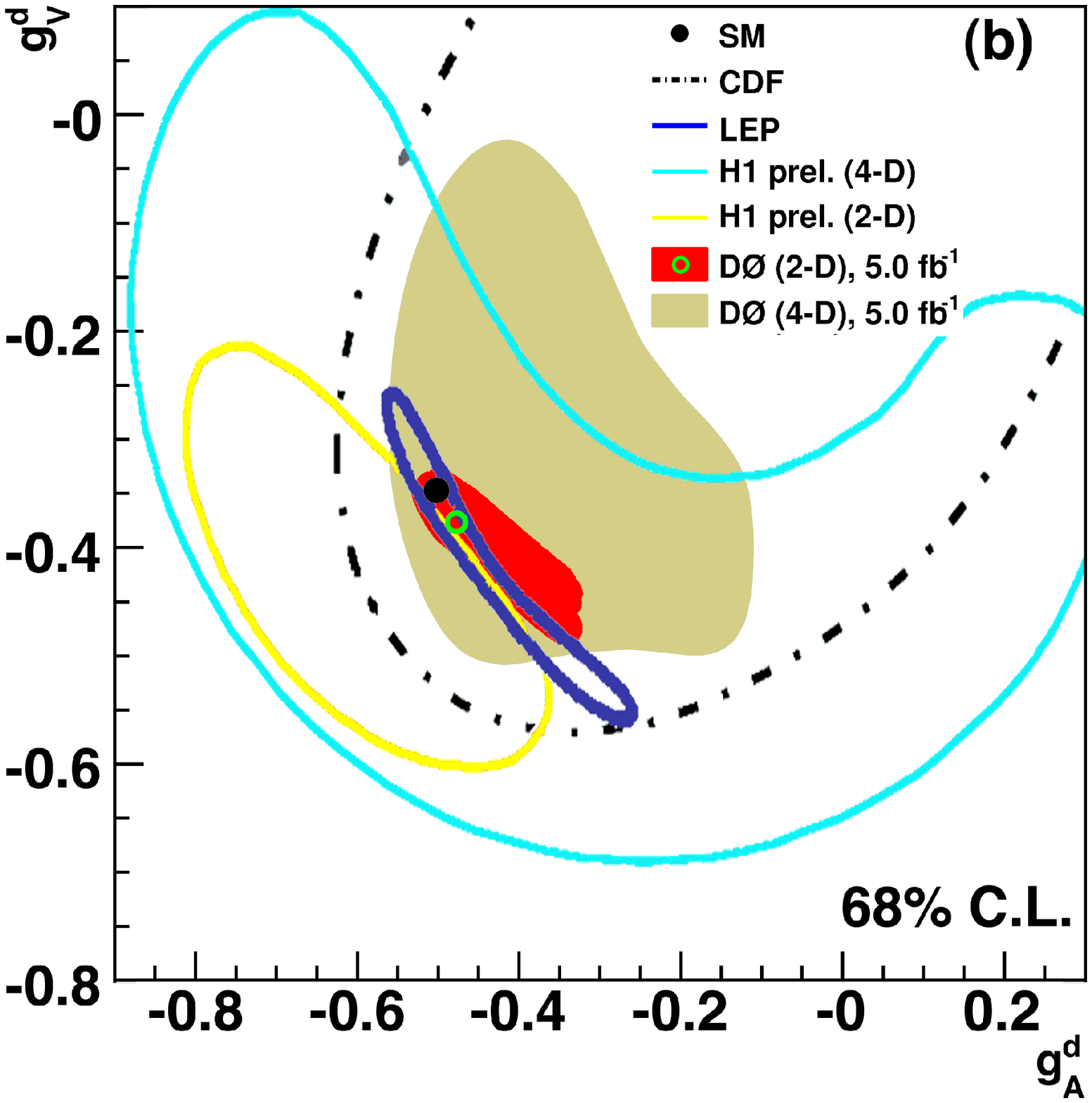}
  \caption{The 68\% C.L. contours of the measured vector and axial-vector couplings of the (a) up-quark and (b) down-quark to the $Z$ boson compared with prior measurements.
    The D\O\ data contours are presented for both a two-dimensional fit, fixing the $u$ ($d$) couplings to the Standard Model values while fitting the $d$ ($u$) couplings,
    and a simultaneous four-dimensional fit.}
  \label{fig:ZuZd}
  \end{center}
\end{figure}
By comparing the unfolded $A_{FB}$ spectrum with templates generated using \textsc{ResBos} for a variety of Z-light quark couplings
the individual measured quark couplings can be determined. For this measurement, the mixing angle is fixed to the global fit value of 0.23153.
The results are summarised in Figure~\ref{fig:ZuZd} for a two and four-dimensional fit procedure. These coupling extractions are the most precise direct measurements to-date.

\section{{\boldmath $W+$}jets production measurements}

\begin{figure}[htbp]
  \begin{center}
  \includegraphics[width=0.43\textwidth]{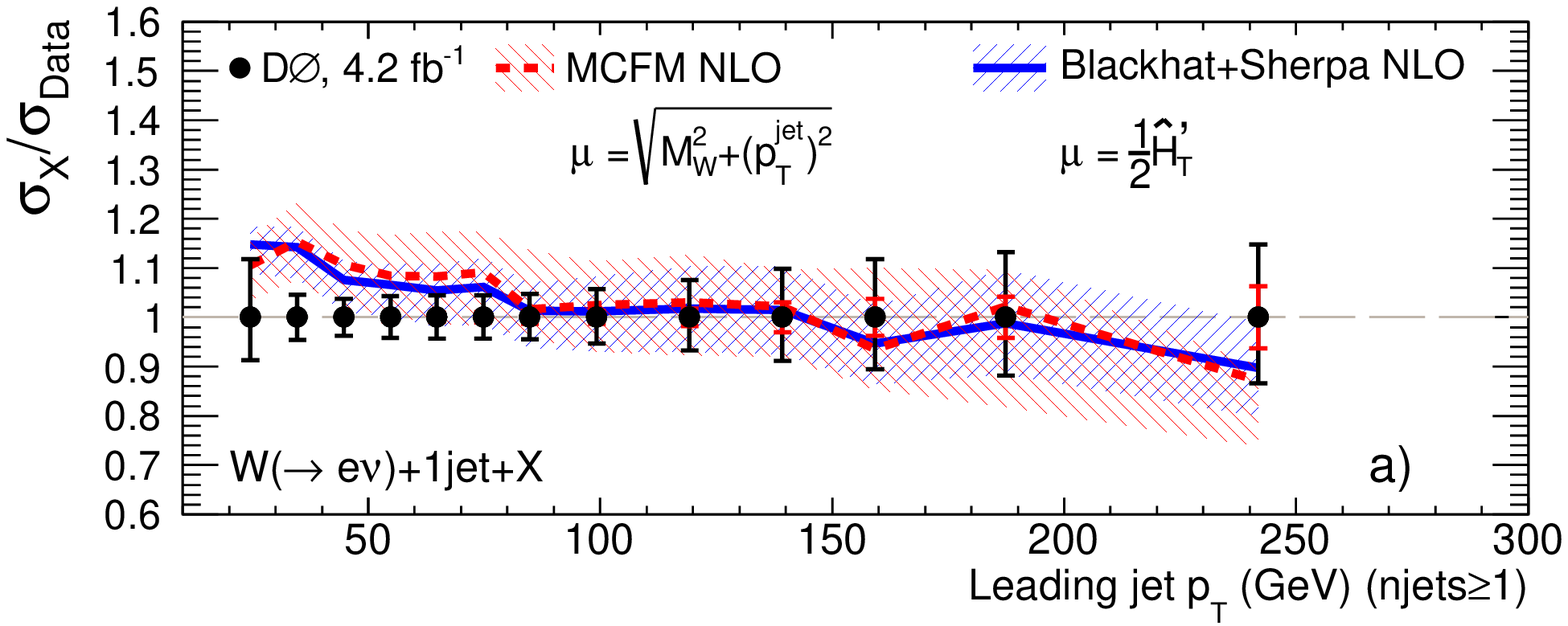}
  \includegraphics[width=0.43\textwidth]{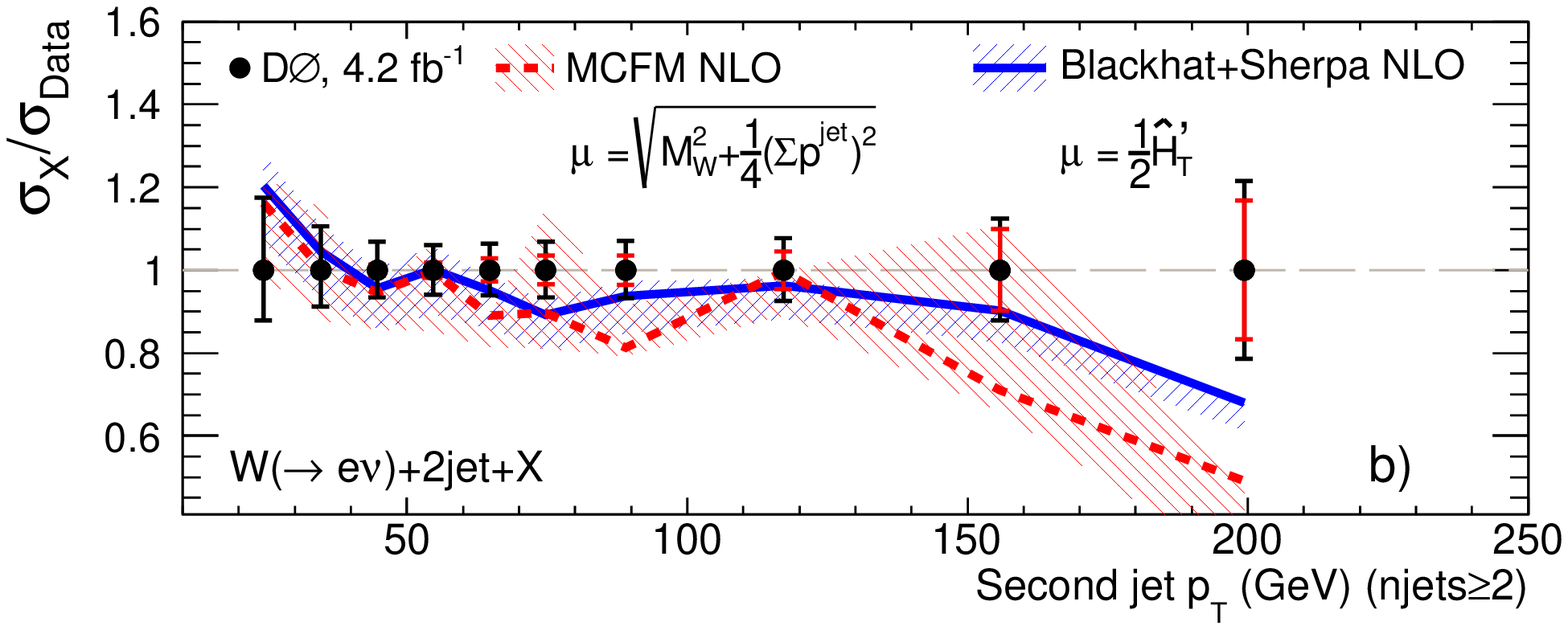}
  \includegraphics[width=0.43\textwidth]{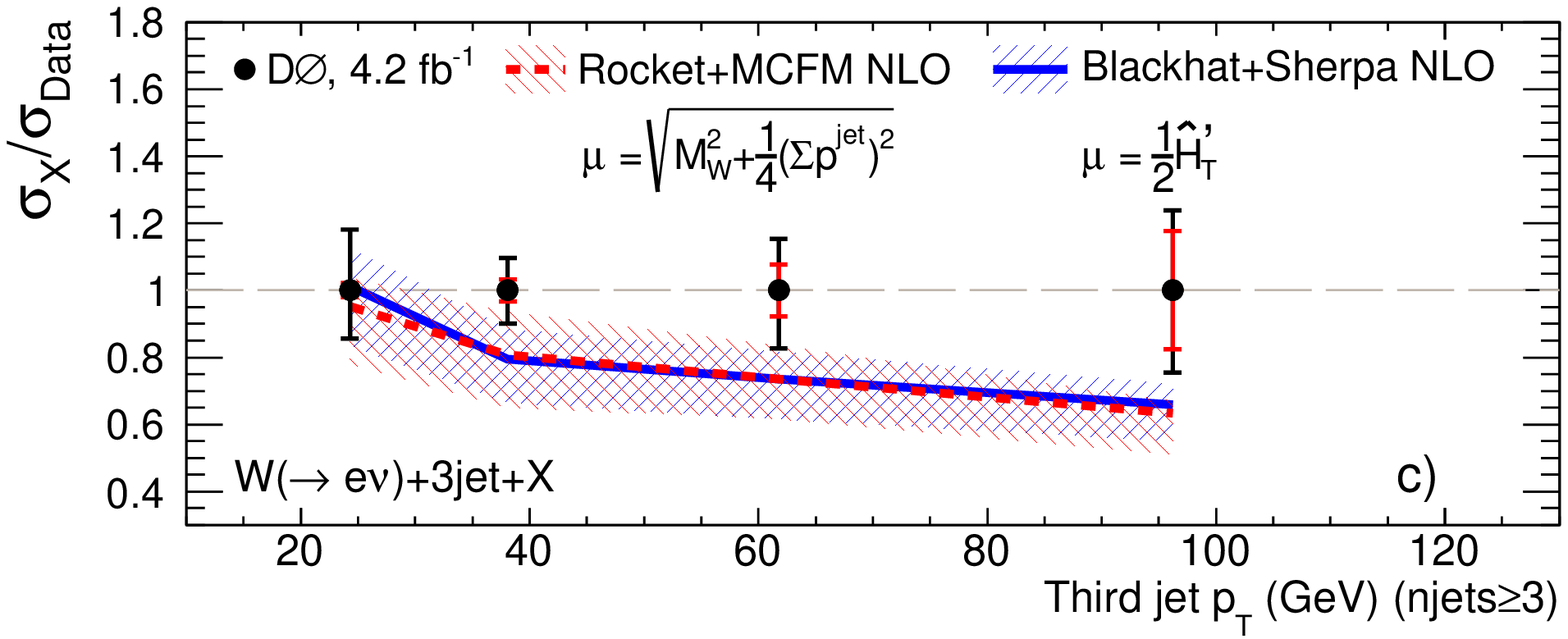}
  \includegraphics[width=0.43\textwidth]{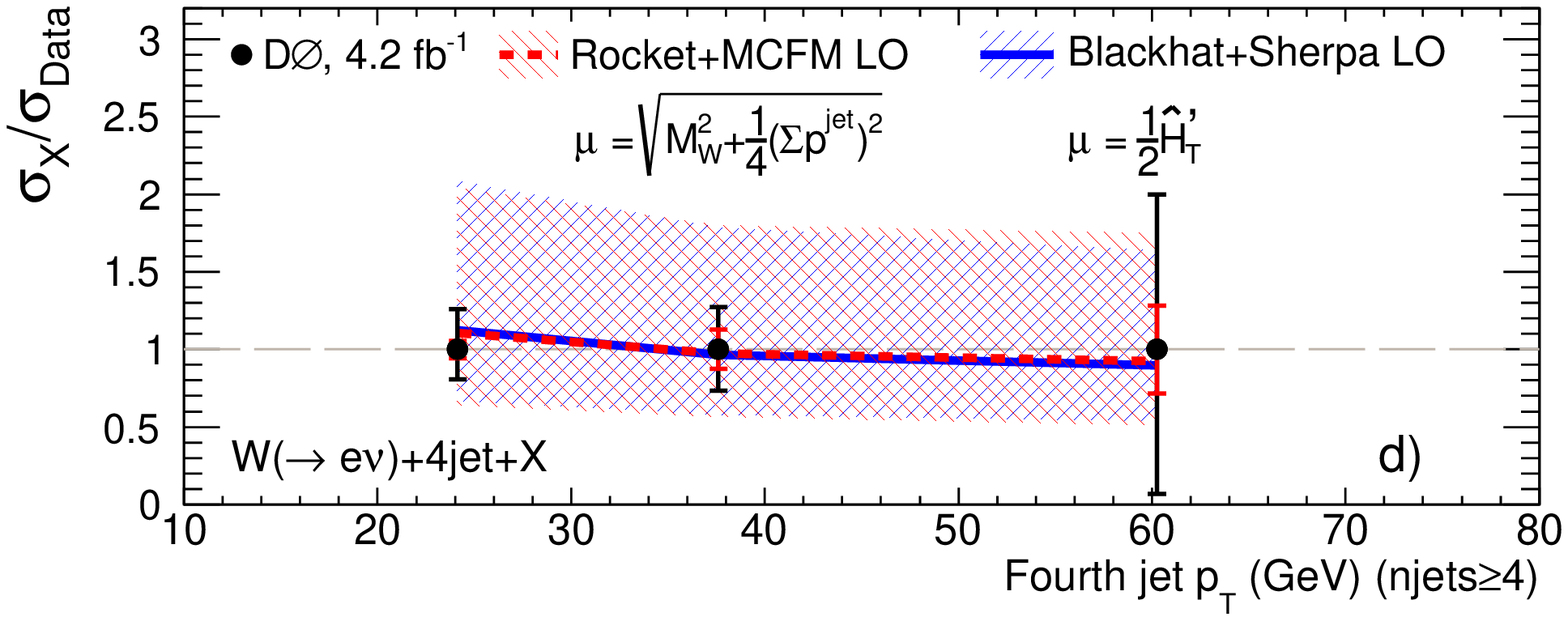}
  \caption{The ratio of pQCD predictions to the measured differential cross sections for the $n^{\mathrm{th}}$ jet $p_T$ ($n=1-4$).
    The corrected data and theory predictions are normalised by the measured inclusive $W$ boson cross section and the predicted inclusive $W$ boson cross sections, respectively.
    The inner (red) bars represent the statistical uncertainties of the measurement, while the outer (black) bars represent the statistical and systematic uncertainties added in quadrature.  
    The shaded areas indicate theoretical uncertainties due to variations of the factorisation/renormalisation scales. 
}
  \label{fig:wjets}
  \end{center}
\end{figure}

The production of a $W$ boson in association with jets was recently studied\,\cite{D0wjets} by the D\O\ Collaboration with 4.2~fb$^{-1}$ of data.
Measurements of the inclusive $W+n$-jet production cross sections and $R=\sigma_{n}/\sigma_{n-1}$ ($n=0-4$) were produced
and differential cross sections as a function of the $n^\mathrm{th}$ jet $p_T$ were unfolded (shown in Figure~\ref{fig:wjets} as the ratio of theory over data). 
These measurements were compared to NLO pQCD calculations (LO for 4-jet).
Reasonable agreement was seen between between unfolded data and theoretical predictions, although some tension was observed in the scaling behaviour between the two
and three jet cross sections in particular.

The CDF Collaboration has measured the production of $W$ bosons in association with $b$-jets\,\cite{CDFwbjet} with 1.9~fb$^{-1}$ of data
and determined the cross section times branching fraction of $W+b$-jets to be $2.74 \pm 0.27~\textrm{(stat.)} \pm 0.42~\textrm{(syst.)}$~pb in significant disagreement
with the prediction from MCFM of $1.22 \pm 0.14$~pb. 

This discrepancy is only apparent in $b$-jets however, as production of $W$ bosons in association with charm jets
was recently studied\,\cite{CDFwcjet} by CDF on 4.3~fb$^{-1}$ of data and the cross section times branching fraction measured to be 
$13.3 ^{+3.3}_{-2.9}~\textrm{(stat. + syst.)}$~pb in comparison to NLO predictions of $11.3 \pm 2.2$~pb.

\section{{\boldmath $Z+$}jets production measurements}

\begin{figure*}[htbp]
  \begin{center}
  \includegraphics[width=0.9\textwidth]{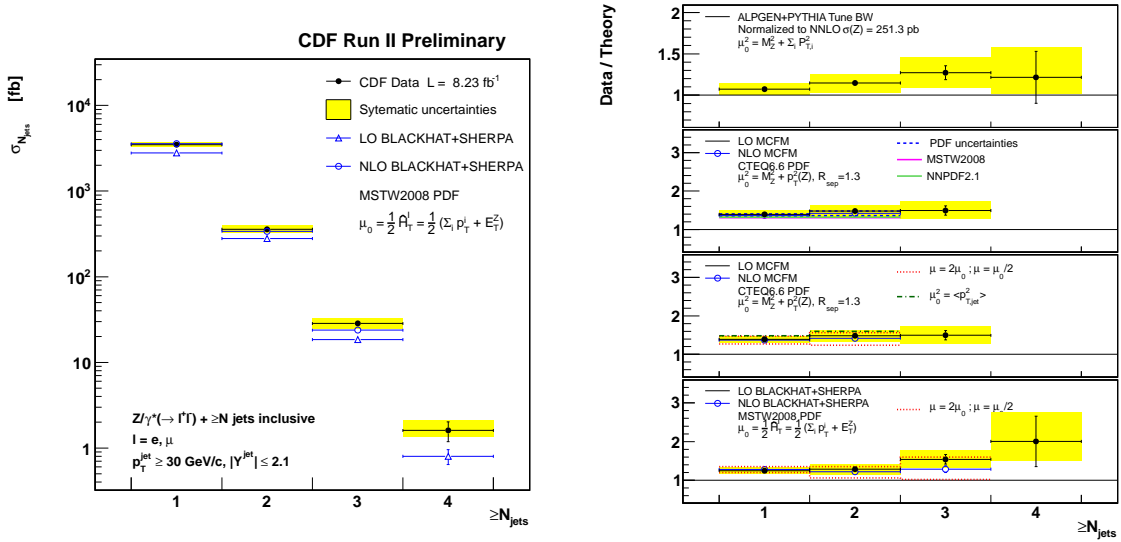}
  \caption{Measurement of the inclusive production cross sections of $Z+(n)$jets for $n=1-4$. Comparison is made to LO/NLO pQCD predictions from \textsc{Blackhat+Sherpa}
and also presented as a ratio of data/theory in comparison with \textsc{Alpgen+Pythia} predictions and LO/NLO pQCD predictions from \textsc{mcfm} for a variety of choices
of scale and parton density functions.}
  \label{fig:zjets}
  \end{center}
\end{figure*}

CDF have recently released detailed inclusive and differential measurements\,\cite{CDFzjets} of the kinematics of $Z$ bosons and jets in $Z+$jet associated production, for up to three jets.
Figure~\ref{fig:zjets} shows the inclusive cross sections for $Z+(n)$jets for $n=1-4$ along with comparisons to LO/NLO pQCD predictions for a variety of scale choices
and parton density functions (pdfs), as well as comparison to the matrix element-parton shower matched \textsc{Alpgen+Pythia} MC predictions. 

Detailed corrected data comparisons to NLO pQCD predictions have been made differentially to a range of kinematic variables\,\cite{CDFzjets}, 
including the inclusive and $n^\mathrm{th}$ jet $p_T$ in $Z+(n=1-3)$-jet events, inclusive jet rapidities, 
$H^\mathrm{jet}_T$, $M_{jj}$, $M_{Zjj}$, $\Delta R_{jj}$, $\Delta\phi_{jj}$, and dijet $p_T$. 

Broadly NLO predictions are found to describe the data well, although several areas are observed where descriptions could be improved.
The \textsc{Blackhat} pQCD prediction differs from \textsc{MCFM} in the choice of scale and some improvement with the \textsc{Blackhat} scale choice 
is observed in the description of the high $p_T$ tail of jet $p_T$ in $Z$+1jet events.
\textsc{Alpgen} gives a good general agreement within the uncertainties. The use of a new $\alpha_s$-consistent tuning of \textsc{Alpgen+Pythia}, Perugia 2011,
leads to improved agreement.
Uncertainties on parton density functions are quite small with respect to the scale uncertainty and in general the measurement cannot be used to distinguish between different pdfs.

The production of $Z$ bosons has also been studied in association with $b$-jets, as for the $W$, both by the D\O\ Collaboration on 4.2~fb$^{-1}$ of data, and the CDF
Collaboration with 7.86~fb$^{-1}$
The D\O\ analysis performed measurement\,\cite{D0zbjet} of the $(Z+b)$/$(Z+$jet$)$ production cross-section ratio, combining not just the secondary vertex invariant mass,
but additional discriminating information such as B-lifetime and decay length significance as inputs to a neural network in order to build discriminant template shapes
for light, charm and beauty jets. These templates were then fit to the data to extract the $(Z+b)$/$(Z+$jet$)$ fraction
$1.92 \pm 0.22~\textrm{(stat.)} \pm 0.15~\textrm{(syst.)}\%$, the most precise measurement of this quantity to-date.

\begin{figure}[htbp] 
  \begin{center}
 \includegraphics[width=0.5\textwidth]{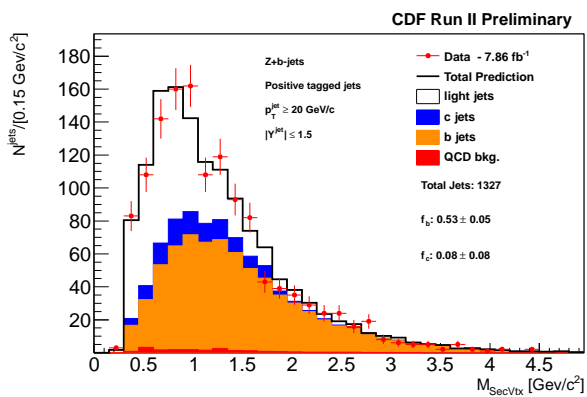}
  \caption{Secondary vertex invariant mass distribution for $Z\to\ell^+\ell^-$ events, used to determine the $Z+b$-jet fractions from data. Fitted \textsc{Alpgen}-derived
    templates for the light, charm and beauty jet contributions to the data are also displayed.}
  \label{fig:zbfraction}
  \end{center}
\end{figure}

The new CDF analysis\,\cite{CDFzbjet} performs a similar discriminant template fit to the data (shown in Figure~\ref{fig:zbfraction}) to extract both the $(Z+b)$/$(Z+$jet$)$ fraction:
$2.24 \pm 0.23~\textrm{(stat.)}$ $\pm 0.32~\textrm{(syst.)}\%$ (in a different phase-space to the D\O\ analysis), and also the $(Z+b)$/$Z$ ratio:
$0.293 \pm 0.030~\textrm{(stat.)}$ $\pm 0.036~\textrm{(syst.)}\%$, both in good agreement with MCFM NLO predictions of $1.8-2.2\%$ and $0.23-0.28\%$ respectively
(range of predictions from different scale choices for the MCFM prediction).

\begin{figure}[htbp]
  \begin{center}
  \includegraphics[width=0.55\textwidth]{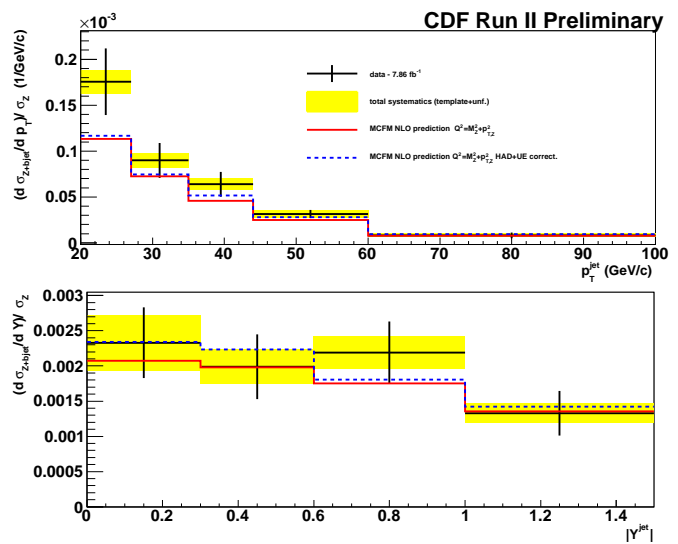}
  \caption{Normalised differential cross section of $Z+b$-jet events as function of jet transverse momentum (top) and as function of jet rapidity (bottom).
  Comparison is made to MCFM NLO predictions with and without non-perturbative corrections applied.}
  \label{fig:zbdiff}
  \end{center}
\end{figure}

The large dataset analysed has also allowed for measurement of the $(Z+b)$/$Z$ ratio to be conducted in bins of jet $p_T$ and rapidity for the first time. 
Doing so leads to a normalised differential production cross-section measurement for $Z+b$-jets as a function of the leading $b$-jet $p_T$ and rapidity, as shown in Figure~\ref{fig:zbdiff}.
This measurement is statistically limited, with statistical uncertainties of around 20\%. Comparison is made to NLO MCFM predictions, with and without corrections
to the theory for non-perturbative hadronisation and underlying event effects. The data show good agreement with the theory predictions, albeit with large uncertainties
on the measurement.

\end{document}